# A User-Friendly Environment for Battery Data Science

Robert Masse[1,2], Dan Ulery[1], Hardik Kamdar[1]

[1]Astrolabe Analytics, Inc.
[2]Materials Science and Engineering, University of Washington

**Abstract**

We report a user-friendly software environment for battery data science. It is designed to streamline data management, data cleaning, and data analysis to help bridge the gap between the domain expertise of most battery scientists and the tools needed as the field becomes increasingly data intensive. The software solution suitable for ingesting battery test data from disparate sources. By aggregating data in an intelligent way, users can streamline routine data analysis tasks and leverage Jupyter Notebook functionality to build advanced scripts and analytics, thereby making battery engineering teams more productive.

**1. Introduction**

It is well understood that electrochemical energy storage using batteries represents a space prime for innovation. However, one of several bottlenecks for the commercialization of new battery technology is at the level of battery research and development. Teams of battery scientists, engineers, and technicians are faced with an array of data management challenges as they test battery performance. Labs may test hundreds or thousands of cells in parallel, with each cell generating megabytes or gigabytes of data, depending on the nature and duration of the test.

As a given battery lab grows, routine data analysis becomes onerous. Different pieces of hardware are accompanied by proprietary software and mutually incompatible data formats that does not scale well over time. Furthermore, while R&D teams are composed of chemists, material scientists, or other engineers, expertise around software development, IT, or data science often falls outside a team's core competencies. As a result, battery engineering teams often lose 25% of their time on rote data management and analysis chores using fragile legacy software solutions. This problem is extant across all stages of the battery value chain - from when raw materials are refined to assembly into electrodes and other subcomponents, all the way through field deployment and commissioning.

This lack of standardization slows down individual researchers, their team's project timelines, and any collaboration between groups. While MS Excel VBA macros are still commonplace, a growing body of software tools and solutions have been developed that each address different aspects of this problem space. This includes, inter alia, commercial software from Voltaiq,[1] the Battery evaluation and early prediction software package (BEEP),[2] the Dahn research group's Universal Battery Database,[3] *cellpy* developed by researchers at IFE Norway,[4] Battery Archive,[5] and Galvanalyser.[6] Preliminary concepts for a battery data genome have also been proposed to help develop best practices for working with and sharing data.[7]

Here we report a suite of workflow tools for battery researchers to lay the foundation for future battery infrastructure and innovation. The platform integrates with major hardware vendors to extract data from installed hardware and load it into a cloud database. A second effort incorporates analytics for routine data visualization, dashboarding and reporting using a web application to save time and provide greater visibility for battery data stakeholders into their testing data. Third, we employ Jupyter Notebooks[8] as a web-based API for custom scripting and an environment for battery data science. The platform promises to accelerate battery R&D and shave significant time off product development efforts.

## 2. Technical Details

This platform is designed to simplify data management for companies across the battery value chain:

1. It streamlines data cleaning, the most tedious in any data analysis project. By integrating our cloud-based software with the hardware installed on-site, we automatically port raw test data into a unified and useful format for analysis.

2. It provides advanced data visualization and analytics tools. Hours of analysis can be done in minutes or seconds to generate standard plots and statistics for cell capacity, voltage profiles, as well as more involved analyses like incremental capacity[9] or other user-defined functions. Templates for standard battery performance metrics can be generated with a few clicks, and more advanced analyses that were previously time-prohibitive are now built-in functions. This is particularly important for battery developers where thousands of battery cells may be tested simultaneously, each being cycled thousands of times.

3. It provides an API for custom analysis and reporting. Our reporting tools provide templates that automatically generate reports that are meaningful to executives, investors, and other non-scientists.

*2.1. Data ingestion*

The breadth of different hardware vendors and file formats for battery testers and potentiostats leaves plenty of room for standardization. The first task of the system is to take in data from disparate sources and collate it into a common data structure. To date, the set of supported hardware and file types are shown in Table 1.

**Table 1.** The platform currently supports datasets from the following hardware vendors in the specified file formats. *Note that many vendors have many different software versions that have been deployed over time, and not every such version may be supported yet.

| Hardware Vendor | Supported File Formats* |
|---|---|
| Admiral Instruments | Excel, .txt |
| Arbin Instruments | .csv, Excel, .res |
| Basytec | .txt |
| Biologic | .mpr, .mpt |
| Land/Landt | Excel |
| Maccor | .csv, Excel, .001 (CSV with channel as file extension) |

| Neware | Excel, .nda |
| Novonix | .csv |
| Versastat | .par |

Data may be uploaded in two ways: a) by uploading discrete sets of files using the web interface (Figure 1), or 2) by downloading a tool to automatically monitor a given local directory and uploading the test data in that folder (Figure 2). The automatic process can be set to ingest new data on a schedule specified by the user in the case of tests that are actively running (e.g. once every morning at 5am). This client-side application uses the .NET WinForms library.

**Figure 1.** Example screenshot of three Arbin .res files being uploaded.

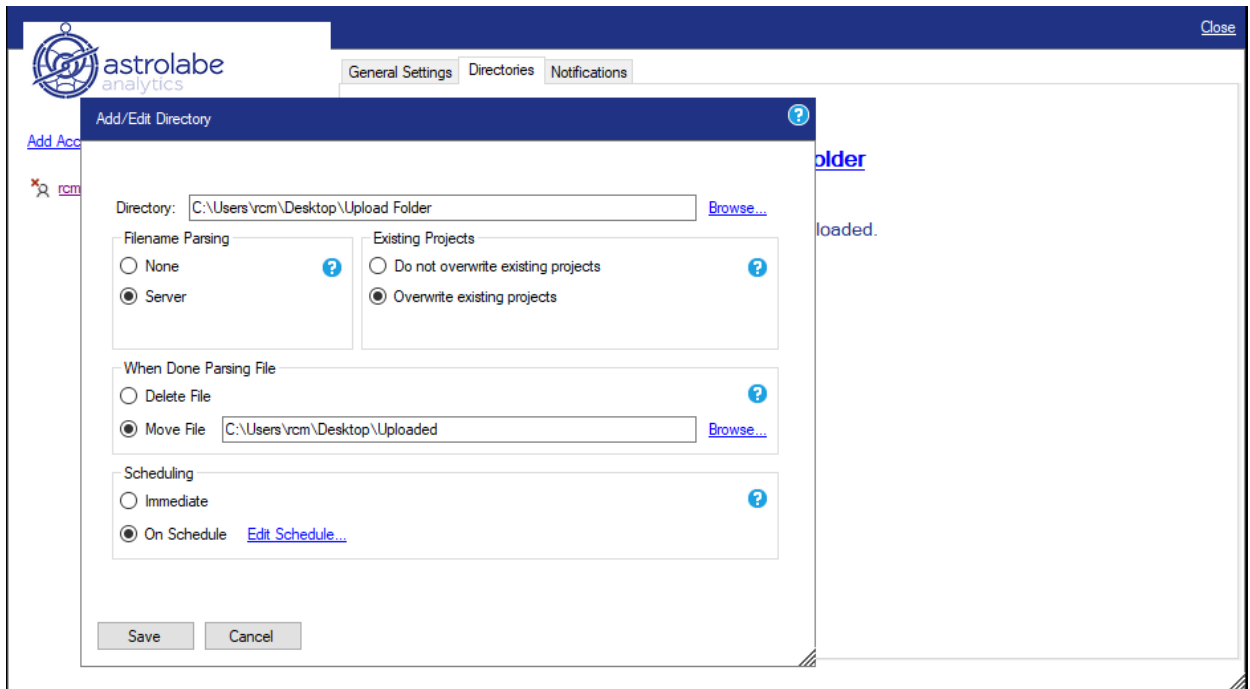

**Figure 2.** User interface for the automatic upload process.

*2.2. Database Server*

The next component of the system is the database, which is SQL Server 2017. The actual files that are uploaded get stored using Amazon Web Services' Simple Storage Service (AWS S3). A configurable retention period is used to delete previous versions of files older than 365 days to avoid using up S3 space.

Web browsers accessing the web app form another client-server connection. The web app runs Angular 11 in the browser and ASP.NET on the server. The server exposes most of the data and functionality using either a RESTful API or via OData. The web app connects to the database server over SSL.

Sharding has been implemented to enable scale out, which plays a role in how the data is structured. For battery testing data, the master node in the sharding implementation holds the following primary tables:

***Projects*:** Holds project metadata. A row in this table represents the data for one test channel. Some file formats contain multiple channels for a single file, so there will be multiple rows created in those situations. Most metadata can be modified by the user via the web app. Examples of the metadata that can be stored here include the filename, file size, number of cycles, active material loading, and other relevant data. See Appendix for further details.

*ProjectTags*: Allows the users to attach their own arbitrary additional metadata to the project. This is essentially a key/value table that has a relation to the project row that it is connected to. Columns are:

> Id
> Name
> Value
> ProjectId

The shard nodes contain the actual data. There are only three tables in our shards:

*DataPoints*: This is the raw time series data. Based on how different hardware handles data logging and acquisition, not all file formats have data for every column in this table, although we calculate those values when possible. Columns include:

> Id
> Capacity
> Current
> CycleIndex
> CycleStep
> Energy
> Index
> Power
> ProjectId
> Temperature
> Time
> Voltage
> StepIndex
> WallTime
> Resistance

*Cycles*: This table contains pre-computed cycle statistics generated from the DataPoint data parsed from the file. This is generated as part of file parsing. Columns include such data as (dis)charge capacity, energy, and power and their statistics (e.g. maximum or minimum power).

*2.3. Web Application*

Detailing every feature of the web app is beyond the scope of this report, but there are a few key operations. Specific usage examples will be given in the next section on Example Use Cases.

*Upload:* As mentioned above (Figure 1), this is the intake of data. The user selects one or more files for upload. The files must have an expected file extension (Table 1), and a Name is a required field. Other metadata fields are available for entry as well as ability to specify permissions. Files are split into chunks to accommodate large file uploads. These chunks are uploaded to a REST endpoint. Once all chunks are uploaded, each file gets a job assigned to it to parse the data and store in the database. Jobs are executed with a tool called Hangfire.

*Project List:* This shows a list of projects the user has uploaded and available. The list is retrieved from various OData endpoints depending on what list is being viewed. An example of this GUI is shown in Figure 3. Items in black have been uploaded and processed, whereas grey text means that it is currently being processed, and red text means that an error has occurred (for instance in the case of data that does not fit a recognized format).

![Figure 3 screenshot of Astrolabe homepage]

**Figure 3.** Screenshot of the homepage for selecting data for further analysis.

*Plotting***:** Once one or more projects are selected, they can be plotted. The basic built-in functionality allows users to select an independent variable for the x-axis from a dropdown menu (e.g. cycle number, or capacity, or time), and then select up to two dependent variables for the y-axes (e.g. for plotting capacity and coulombic efficiency vs. cycle number, Figure 4).

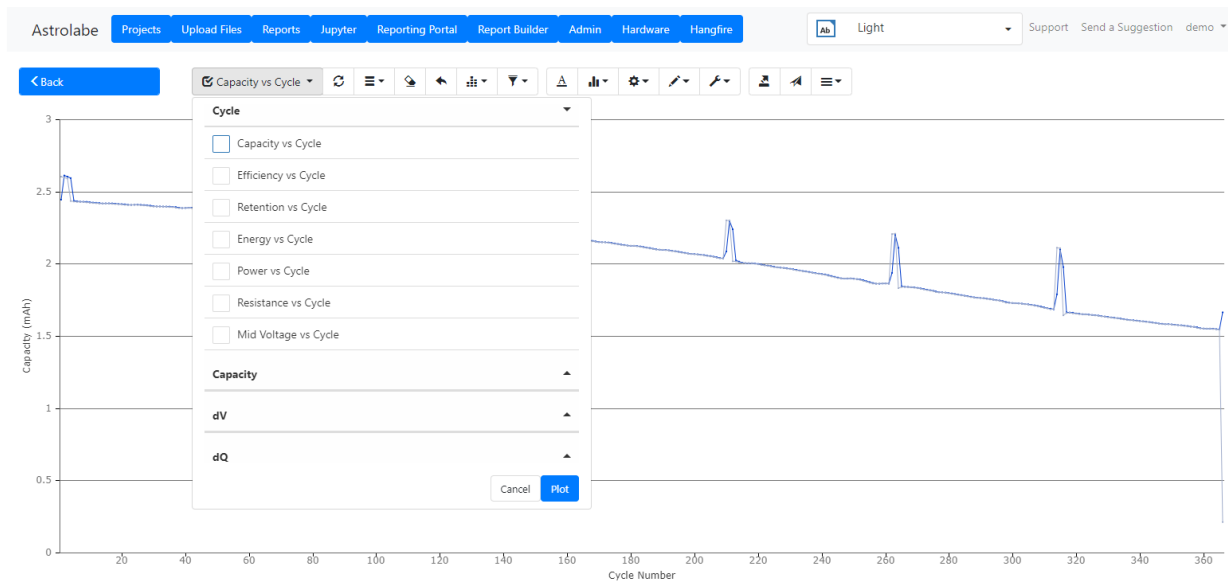

**Figure 4.** Example screenshot plotting cycling statistics.

The client sends the list of projects along with information about which plot is requested as well as parameters for the plot. Other parameters may have already been stored as defaults on the server side. Using all these parameters, the server sends back data points that are then plotted using CanvasJS.

It is worth noting that many different projects are used to provide functionality in the web app, although the client side uses DevExtreme GUI components and CanvasJS for chart generation.

Users can belong (optionally) to one or more Organizations, which are essentially groups of users. Users may assign projects to an Organization for sharing across a given team, or from their organization to a downstream data stakeholder.

*2.4. Jupyter Notebook*

Jupyter Notebook comes equipped with kernels for several different languages without installing anything. This allows power users to put together custom scripts and analytics for use cases that are not built-in to the core webapp. Users can query the database using the `project_search` library. These queries return Pandas DataFrames (Figure 5). This API also keeps our database structure, and more specifically the querying of data cross database shards agnostic. That is, users can specify certain projects and then additionally specify whether or not to include time series data (`query.includeDataPointData()`), cycle statistics (`query.includeCycles()`), and/or other custom metadata about a given project `query.includeProjectTags()`).

Users will find the standard set of scientific computing libraries preinstalled (such as NumPy,[10] SciPy,[11] Matplotlib,[12] scikit-learn,[13] and pandas[14]), as well as libraries for advanced visualizations (Altair[15]), and battery-specific tools such as cellpy[4] and Pybamm.[16] We are interested in collaborating with the wider community to establish best practices and interoperable tools and solutions for data pipelining, data-driven modeling and other applications based on this work to date.

Jupyter (via Jupyterhub functionality) is distinct from web app functionality. The only link there is that the web app is used to provide authentication information to Jupyter, specifically, to which Organizations a user belongs.

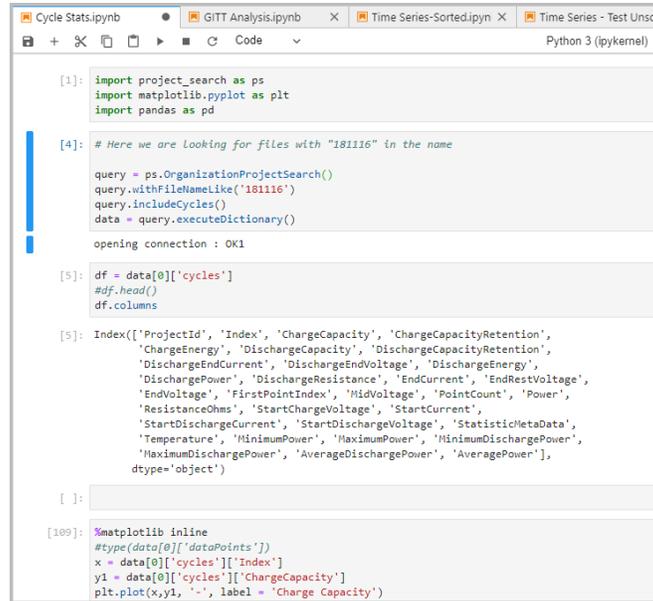

**Figure 5.** Screenshot of Jupyter Notebook interface and sample query using the `project_search` library.

### 3. Example Use Cases

While this is not intended as the full documentation for the software, a few illustrative use cases are presented here.

*3.1. Getting started*

A preliminary video tutorial for new user registration, uploading data, and basic plotting of cycling statistics, voltage profiles, and differential capacity plots is available from Ref. 17.

*3.2. Templates for Generating Voltage Profiles*

For frequently generated plotting, users may build a template that produces a reproducible output for a given plot. For example, users may select Voltage vs Capacity from the Templates dropdown menu and specify an interval or custom range of cycles to generate a re-usable template for voltage profiles (Figure 6). Additional optional settings for chart formatting are also available in this menu.

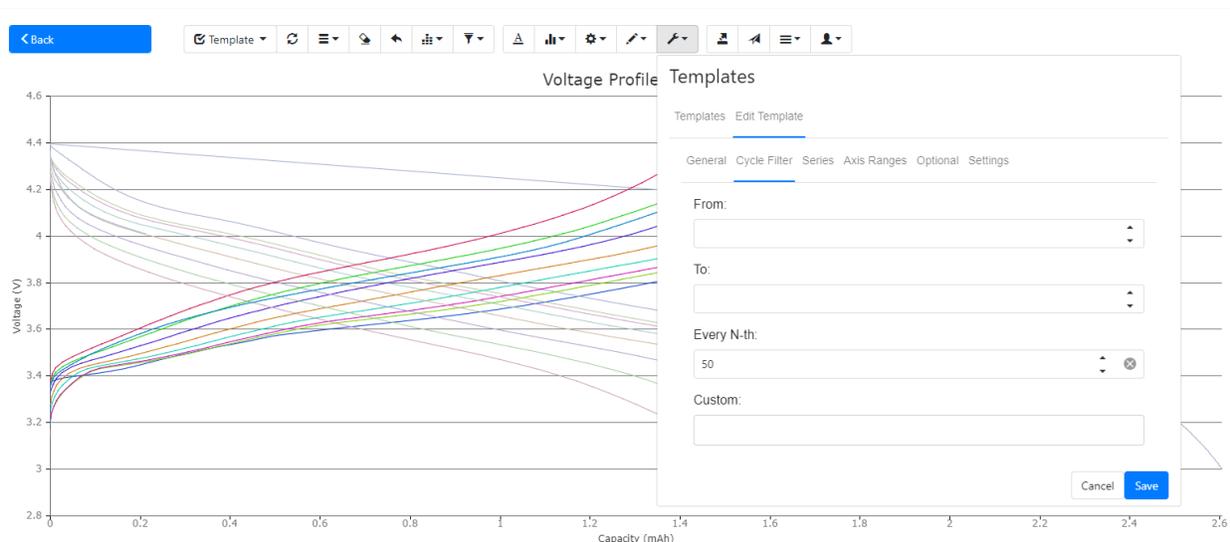

**Figure 6.** Example screenshot of the cycle selector tool used to generate a template for repeated analysis.

*3.3. Differential Capacity Analysis*

Differential capacity or incremental capacity analysis are diagnostics that have been identified as leading indicators of battery degradation.[9,18,19] By taking the numerical derivative of capacity stored (discharged) over a given voltage step size, peaks emerge that correspond to phase transformations that are characteristic to a given battery material chemistry. Over time and with cycling, peak intensity tends to go down as capacity decreases and peak position will change as the internal resistance and electrochemical overpotentials change (Figure 7). As such, this is a useful tool for battery health forecasting. For example, shifts in the potential where Li intercalates into graphite are known to emerge before capacity fade is noticeably affected.[20,21]

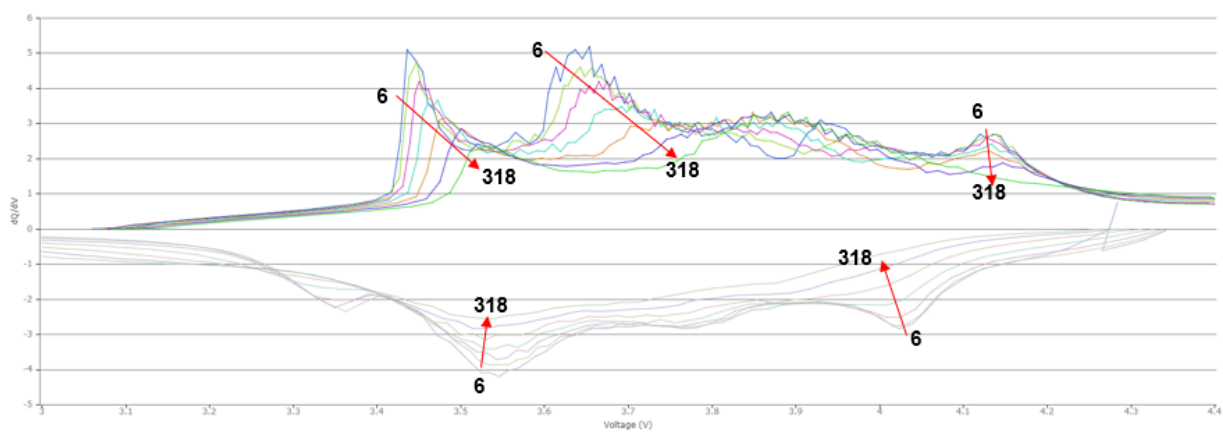

**Figure 7.** Advanced analyses such as differential capacity analysis can be used to resolve the evolution of electrochemical processes over time. Using built-in capabilities already deployed, we find peak position and intensity changes during cycling, suggesting an evolution in overpotential and loss of capacity.

*3.4. Galvanostatic Intermittent Titration Technique (GITT)*

The webapp provides functionality that covers many but not all routine battery data analysis applications. GITT is an example where Jupyter can be deployed to cover those applications not natively supported in the webapp, and give the user the ability to write and run custom scripts. Taking data from a GITT measurement made using a Maccor battery tester, a basic Jupyter Notebook document can be written to run the analysis and print the desired output (Figure 8).

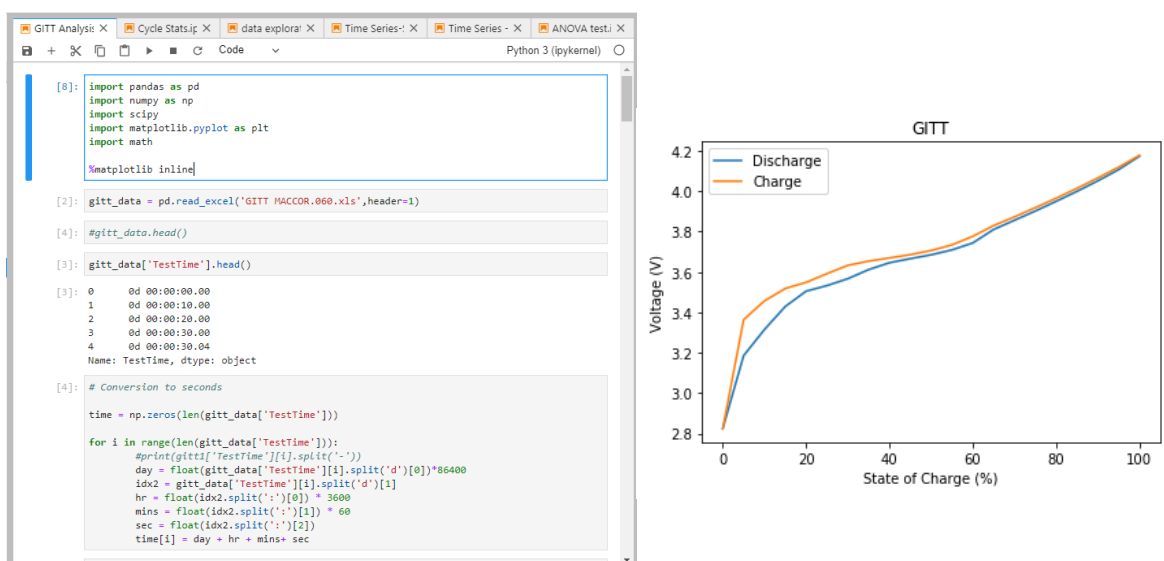

**Figure 8.** a) A portion of the Python script and b) output figure generated for GITT analysis.

## 4. Conclusion

Modern software and data science tools allow for many new opportunities to better equip battery scientists and engineers. While battery modeling researchers are familiar with tools such as MATLAB and Python, many experimentalists do not have this background. Here we described a platform for battery data management and analysis that allows users to streamline data chores and more rapidly iterate on data-intensive projects and hypotheses.

This platform is a web-based application that consists of workflow tools to run the standard analyses familiar to any battery materials scientist or engineer, as well as an API based on Jupyter Notebook for more advanced custom analyses.

Our goal is to help grow an ecosystem of battery innovators that builds on the similar work of others in this community, with the objective of tightening the feedback loop between basic science at academic institutions and applied engineering that takes pace in industry to advance battery technology. We are eager to collaborate with others in the battery community to build on this work and help develop standards for battery data best practices.

**Appendix**

For reference, the *Projects* table referenced in Chapter 6.2.2. has the following columns:

    Id
    ActiveMaterialFraction
    Area
    Channel
    Comments
    CreatedAt
    Error
    Failed
    FileName
    FileSize
    InternalFileName
    IsAveragePlot
    IsPartialGathering
    IsReady
    JobId
    Mass
    Name
    StitchedFrom
    StitchedFromNames

        Tag
        TestName
        TestType
        TheoreticalCapacity
        TraceId
        UpdatedAt
        UserId
        NumCycles
        TestDate
        PAMMass
        NAMMass
        IsRealTime
        ProcessDate
        IsProcessing
        ProcessingMessage
        DataPointStartDate
        Shard_Id
        Organization_OrganizationId
        ExtraDataNameJSON
        ErrorDetailed

*DataPointExtraData*: Some formats contain auxiliary data. An example is VARx, FLGx columns in Maccor. This is where we store that data. There could be multiple aux data values for each data point (i.e. VARx1, VARx2, etc.). In order to save on database space, we only store aux data on the data point in which the value has changed compared to the previous value for this aux data. Columns are:

        Id
        DataPoint_Id
        Name
        Value

Columns for the *DataPoints* table include:

        Id
        Capacity
        Current
        CycleIndex
        CycleStep
        Energy
        Index
        Power
        ProjectId
        Temperature

Time
Voltage
StepIndex
WallTime
Resistance

Columns for the *Cycles* table include:

ProjectId
Index
ChargeCapacity
ChargeCapacityRetention
ChargeEnergy
DischargeCapacity
DischargeCapacityRetention
DischargeEndCurrent
DischargeEndVoltage
DischargeEnergy
DischargePower
DischargeResistance
EndCurrent
EndRestVoltage
EndVoltage
FirstPointIndex
MidVoltage
PointCount
Power
ResistanceOhms
StartChargeVoltage
StartCurrent
StartDischargeCurrent
StartDischargeVoltage
StatisticMetaData
Temperature
MinimumPower
MaximumPower
MinimumDischargePower
MaximumDischargePower
AverageDischargePower
AveragePower

An additional note on StatisticMetaData: This is JSON text that contains some additional data. There are a number of columns in this JSON. They are:

ChargeCapacityAverage
ChargeCapacityFirst
ChargeCapacityLast

ChargeCapacityMax
ChargeCapacityMin
ChargeCapacityRetentionStdDev
ChargeCapacityStdDev
ChargeCapacityStdError
ChargeCapacityVariance
ChargeEnergyStdDev
ChargeVoltageAverage
ChargeVoltageMax
ChargeVoltageMin
ChargeVoltageStdDev
ChargeVoltageStdError
ChargeVoltageVariance
CoulombicEfficiencyAverage
CoulombicEfficiencyStdDev
DischargeCapacityAverage
DischargeCapacityFirst
DischargeCapacityLast
DischargeCapacityMax
DischargeCapacityMin
DischargeCapacityRetentionStdDev
DischargeCapacityStdDev
DischargeCapacityStdError
DischargeCapacityVariance
DischargeEndCurrentStdDev
DischargeEndVoltageStdDev
DischargeEnergyStdDev
DischargePowerStdDev
DischargeResistanceStdDev
DischargeVoltageAverage
DischargeVoltageMax
DischargeVoltageMin
DischargeVoltageStdDev
DischargeVoltageStdError
DischargeVoltageVariance
EndCurrentStdDev
EndVoltageStdDev
MidVoltageStdDev
PowerStdDev
ResistanceOhmsStdDev
VoltageAverage
VoltageFirst
VoltageLast
VoltageMax
VoltageMin
VoltageStdDev

VoltageStdError
VoltageVariance